\documentclass[referee]{aa}
\usepackage{graphicx}
\usepackage{natbib}
\usepackage{amssymb}
\usepackage{booktabs}

\begin{document}

\title{A HST study of the environment of the Herbig Ae/Be star
       LkH$\alpha$~233 and its bipolar jet
       \thanks{Based on observations made with the NASA/ESA
       {\it Hubble Space Telescope}, obtained at the Space
       Telescope Science Institute, which is operated by the
       Association of Universities for Research in Astronomy, Inc.,
       under NASA contract NAS5-26555.}}

\author{Stanislav Melnikov \inst{1,2} \and
        Jens Woitas \inst{1} \and
        Jochen Eisl\"offel \inst{1} \and
        Francesca Bacciotti \inst{3} \and
        Ugo~Locatelli \inst{4} \and
        Thomas~P.~Ray \inst{5}}

\institute{Th\"uringer Landessternwarte Tautenburg, Sternwarte 5,
           07778 Tautenburg, Germany \and
           Ulugh Beg Astronomical Institute, Astronomical str. 33, 700052
	   Tashkent, Uzbekistan \and 
           I.N.A.F. - Osservatorio Astrofisico di Arcetri, Largo E. Fermi 5,
           50125 Firenze, Italy \and
           Dipartimento di Matematica,
           Universit\`a degli Studi di Roma ``Tor Vergata'',
           Via della Ricerca Scientifica 1, 00133 Roma, Italy \and
           Dublin Institute for Advanced Studies, 5 Merrion Square,
           Dublin 2, Ireland}

\offprints{Stanislav Melnikov,
\email{melnikov@tls-tautenburg.de}}

\date{Received / Accepted}

\abstract{LkH$\alpha$~233 is a Herbig Ae/Be star with a collimated bipolar
  jet. As such, it may be a high-mass analogue to the classical T Tauri stars
  and their outflows.} 
{We investigate optical forbidden lines along the LkH$\alpha$~233 jet to
  determine physical parameters of this jet (electron density
  $n_\mathrm{e}$, hydrogen ionisation fraction $x_\mathrm{e}$, electron
  temperature $T_\mathrm{e}$). The knowledge of these parameters allows us a direct
  comparison of a jet from a Herbig star with those from T Tauri stars.} 
{We present the results of HST/STIS and WFPC2 observations of
  \object{LkH$\alpha$~233} and its environment. These are the first
  observations of this object with a spatial resolution of $\le 0\farcs1$ at
  optical wavelengths. Our STIS data provide spectroscopic maps that allow us
  to reconstruct high angular resolution images of the bipolar jet from
  LkH$\alpha$~233 covering the first $\approx$~2000~AU from the star in the
  blueshifted outflow lobe and  $\approx$~4000~AU in the redshifted
  lobe. These maps are analysed with a diagnostic code that yields
  $n_\mathrm{e}$, $x_\mathrm{e}$, $T_\mathrm{e}$, and mass density $n_\mathrm{H}$
  within the jet.} 
{The WFPC2 images in broad-band filters clearly show a dark
  lane caused either by a circumstellar disk or a dust torus. The
  circumstellar environment of LkH$\alpha$~233 can be interpreted as a conical
  cavity that was cleared by a bipolar jet. In this interpretation, the
  maximum of the optical and near-infrared brightness distribution does not
  coincide with the star itself which is, in fact, deeply extincted. In the
  blueshifted lobe, $n_\mathrm{e}$ is close to or above the critical density for [SII]
  lines ($2.5\times10^4 \mathrm{cm}^{-3}$) in the first arcsecond and
  decreases with distance from the source. The ionisation $x_\mathrm{e} \approx 0.2 -
  0.6$ gently rises for the first 500~AU of the flow and shows two
  re-ionisation events further away from the origin. The electron temperature
  $T_\mathrm{e}$ varies along the flow between $10^4\,\mathrm{K}$ and
  $3\times10^4\,\mathrm{K}$. The $n_\mathrm{H}$ is between $3\times10^3$ and
  $10^5\,\mathrm{cm}^{-3}$, and the mass flux $\dot{M}\approx 10^{-8} -
  10^{-7} \mathrm{M}_{\sun}\mathrm{yr}^{-1}$. The (radial) outflow velocities
  are $\approx 80 - 160 \,\mathrm{km}\,\mathrm{s}^{-1}$, and they appear to 
  increase with distance from the source. In the redshifted lobe, the excitation
  conditions are quite different: $T_\mathrm{e}$, $n_\mathrm{e}$,
  $x_\mathrm{e}$, and $n_\mathrm{H}$ are all lower than in the blueshifted
  lobe, but have the same order of magnitude.} 
{All these derived parameters are just beyond or at the upper limits of those
  observed for classical T Tauri star jets. This may indicate that the flows
  from the higher mass Herbig stars are indeed scaled-up examples of the same
  phenomenon as in T Tauri stars.} \keywords{ISM: jets and outflows -- Stars:
  circumstellar matter -- Stars: emission-line, Be -- Stars: individual:
  LkH$\alpha$~233} 

\titlerunning{A HST study of the Herbig Ae/Be star
              LkH$\alpha$~233}
\maketitle

\section{Introduction}
\label{intro}
Herbig Ae/Be stars (HAeBes) are intermediate-mass young stars, covering the
mass range between 2 and $10~M_{\sun}$. They are less well-studied than their
lower mass counterparts, the T~Tauri stars, for several reasons. First, they
are rare because the initial mass function declines steeply with rising
mass. Second, they spend much less time in their pre-main sequence phase than
low-mass young stars. Third, they are embedded within envelopes, which makes
observations of their immediate environment difficult. 

An interesting question is whether the environment of HAeBes is similar to
that of lower mass young stars, for which the pre-main sequence phases are
relatively well known. In particular one would like to know if mass accretion
from circumstellar disks and mass ejection by means of jets and outflows
(e.\,g. \citealp{Har95,Koe00,Fer02}) is also at play in the formation of these
higher mass stars. This would help in understanding if the accretion/ejection
engine, a key element regulating the formation of low-mass stars, can be
generalised to different stellar masses. Very high-mass young stars are known
to possess disks and outflows \citep[see][]{Zha05}, and recently, also jets
from brown dwarfs have been detected \citep{Whe05,Whe07}. To unveil the nature
of the circumstellar surroundings of intermediate mass stars like  HAeBes
would thus allow us to add an important piece to the star formation puzzle. 

Indeed, some parsec-scale outflows from HAeBes are known \citep{McG04} similar
to those from low-mass stars. To obtain information about the
accretion/ejection engine, however, high angular resolution observations are
necessary. The physical processes involved are confined to a region of at most
$500 - 1000$ AU from the central star. Therefore, even for the nearest star
formation regions, sub-arcsecond resolution is necessary for observations
aimed at testing the accretion/ejection models. Following these ideas, we have
observed the HAeBe star LkH$\alpha$~233 and its environment with the Hubble
Space Telescope (HST), which reaches an angular resolution of $0\farcs1$ in
the optical. 

LkH$\alpha$~233 has been classified as a HAeBe (spectral type A7e) in the
original paper by \citet{Her60} that defined the group. According to
\citet{Hern04}, the star has a mass of $M_*=2.1\,M_{\sun}$, a luminosity of $L_*=21.9
L_{\sun}$, and an age of about 7 Myr. These parameters were calculated with a
normal extincion law (i.e. $R=3.1$ and $A_\mathrm{V}=2\fm3$). The extinction, however, is
quite ambiguous in this region \citep{Hern04} and real parameters may be
different from those. This star is located within a dark cloud that contains
several other young stars. Throughout this paper we will assume the distance
of LkH$\alpha$~233 to be 880~pc and its heliocentric radial velocity to be
$-17.2$~km\,s$^{-1}$. These quantities were measured for \object{HD~213976},
which is a bright B1.5V star associated with the same dark cloud \citep{Cal78,
Wil53}. LkH$\alpha$~233 is embedded within a reflection nebula that shows
spikes at position angles $\mathrm{PA} = 50^{\circ}/230^{\circ}$ and
$90^{\circ}/270^{\circ}$. A first high angular resolution study of the central
source and its close environment has been presented by \citet{Lei93}. Using
speckle interferometry in the near-infrared, they obtained information on a
sub-arcsecond scale and resolved a halo that has a radius of
about~1000~AU. \citet{Cor97} noticed the presence of strong optical forbidden
line emission from LkH$\alpha$~233, indicative of a small-scale jet from this
star. This jet was indeed discovered later by the same authors using longslit
spectroscopy \citep{Cor98}. We observed the small-scale bipolar jet
associated with this source with the Space Telescope Imaging Spectrograph
(STIS) and have taken additional data obtained with the Wide Field Planetary
Camera (WFPC2) from the HST Archive. 

The analysis of these data has led us to determine the structure of the jet at
its base on small scales, allowing a comparison with analogous outflows from T
Tauri stars in the same region of the plasma jet. In addition, the application
of a well-tested spectroscopic technique for jet diagnostics  to the observed
line ratios \citep{Bac99a} has provided values for key physical quantities
along the jet at high spatial resolution, a step that is necessary for 
identifying
the dynamics of the outflow. The obtained results point toward a strong
similarity between the accretion/ejection engine in T Tauri stars and HAeBes,
opening interesting prospects for a generalisation of the mechanism to all
YSOs (Young Stellar Objects).

In our paper, the methods of observations and data analysis are presented in
Sect.\,\ref{obs}, and the results are illustrated in Sect.\,\ref{results}. In
Sect.\,\ref{disc} we discuss the physical implications of our findings and 
compare our results with previous studies of jets powered by lower mass
YSOs. Finally, in Sect.\,\ref{concl} we summarise our conclusions. 

\section{Observations and Data Analysis}
\label{obs}
The jet of LkH$\alpha$~233 was observed with the STIS spectrograph on board
the HST on 4 Oct 1999. The method of data acquisition was virtually identical
to HST/STIS observations of the small-scale jets from the T~Tauri stars
\object{DG~Tau} \citep{Bac00} and \object{RW~Aur} \citep{Woi02}. The
spectrograph slit was moved across the jet in seven steps ($S1$...$S7$) of
0\farcs07 from NW to SE, keeping it parallel to the jet axis (PA =
248$^{\circ}$, \citealt{McG04}). The observations made use of the G750M
grating that, in addition to  H$\alpha$, covers the most prominent forbidden
emission lines [OI]\,$\lambda\lambda 6300, 6363$, [NII]\,$\lambda\lambda 6548,
6583$, and [SII\,]$\lambda\lambda 6716, 6731$. The pixel scales are 0.554
{\AA}/pixel and 0\farcs05/pixel for the dispersion and spatial direction,
respectively. To ensure optimum positioning, the slit was peaked on the source
before offsetting for each observation. As we show in the following,
however, the maximum of the optical brightness distribution does not coincide
with the star. As a consequence, the central slit position, $S4$, does not
coincide with the jet axis. The jet axis is instead closer to $S3$, and $S6$
and $S7$  unfortunately contain no signal from the jet. 

The basic data reduction steps were carried out using the HST/STIS
pipeline. Cosmics and bad pixels were removed from the frames. To see the jet
in the forbidden emission lines, the continuum contribution from the
surrounding reflection nebula had to be subtracted carefully. This was done 
by fitting the continuum around the lines in each pixel row with a
high-order polynomial and then subtracting it from the frames. The spectra
were then used to reconstruct 2D images of the jet in the various lines and to
diagnose the thermal conditions of the gas through our spectral diagnostics
involving line ratios. 

The errors on the line fluxes were estimated considering the contribution of
noise and  uncertainties on the extinction,  $A_\mathrm{V}$. The noise was
estimated from the subtracted background as running averages of boxes of
$7\times8$ pixels in size moved along both sides of the lines. The lines
included in our STIS setting are too close in wavelength to allow a reliable
measurement of the extinction. An estimate in the region around this star is
given, however, by \citet{Hern04} who report a range between  $2\fm3$ and
$3\fm7$. To evaluate the possible effect of reddening, the lines were
dereddened using the median value, and the uncertainty of $A_\mathrm{V}$ was
estimated from the limiting values. The reddening uncertainty was added to the
noise, but turned out to be only a few percent of the latter. Despite
the fact that we could not measure the variation in extinction along the jet
beam, one can therefore expect the resulting flux uncertainty to be 
negligible. Thus our further analysis was done using the measured flux values. 

\begin{figure*}
\centering
\includegraphics[width=6.5cm]{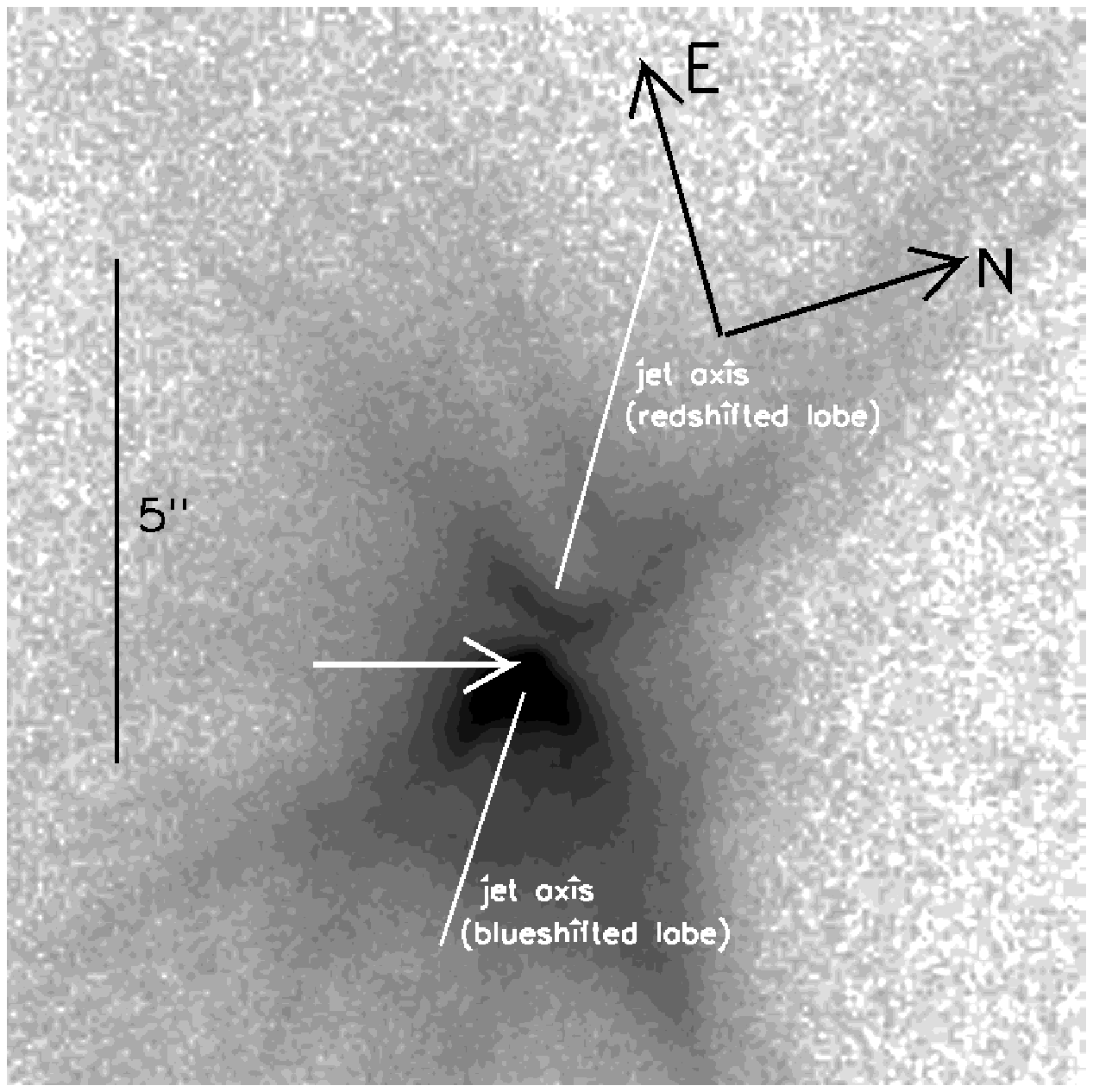}
\includegraphics[width=6.5cm]{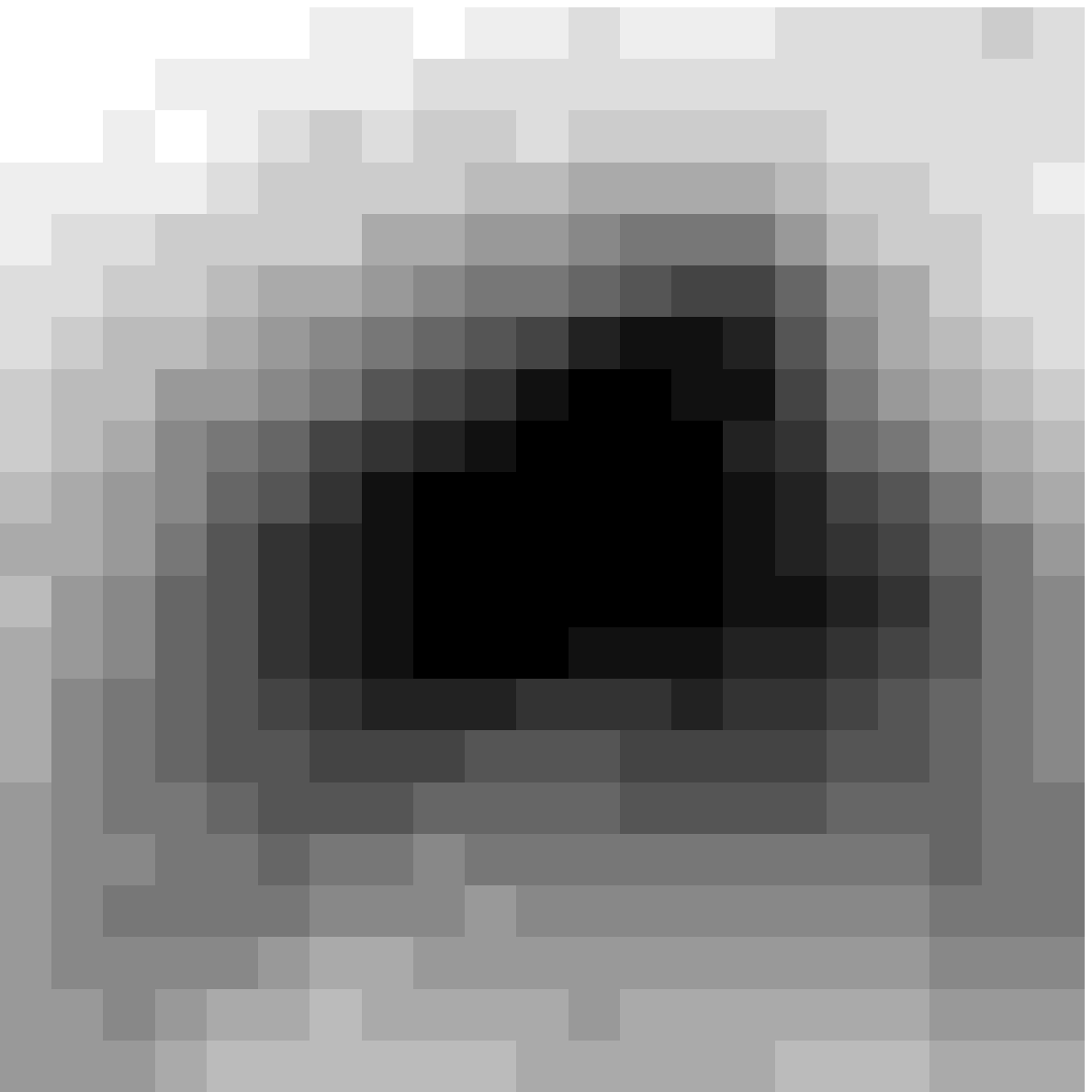}
\caption{\label{lkha233_disk} HST/WFPC2 image of LkH$\alpha$ 233 in the
  broad-band filter W814 (left panel). The axis of the bipolar jet
  (PA~=~248$^{\circ}$ for the blueshifted lobe) is indicated. The jet is not
  visible in this image because the signal is dominated by continuum emission
  from the reflection nebula surrounding the star. The maximum of the
  brightness distribution whose size scale is $0\farcs5 \times 0\farcs 5$
  (marked with the white arrow in the left panel) is magnified in the right
  panel.} 
\end{figure*}

HST/WFPC2 images of LkH$\alpha$~233 were taken on 23 Jan 2001 in the
broad-band filters F606W and F814W (Proposal ID 8216, P.\,I. Karl
Stapelfeldt). After receiving the pipeline-reduced files from the HST Archive,
cosmics and bad pixels were eliminated. Figure\,\ref{lkha233_disk} shows 
images,
with the target on the PC chip of the camera, with a scale of
0\farcs046/pixel. The achieved angular resolution is given by the FWHM of the
PSF that can be calculated using the {\it Tiny Tim} software \citep{Kri95} and
is $\la 0\farcs1$ for the filters used. Previous adaptive optics observations
of LkH$\alpha$ 233 presented by \citet{Per04} have an FWHM of 0\farcs27. 

\section{Results}
\label{results}


\subsection{The reflection nebula}
\label{morph_neb}

The WFPC2 images in both broad-band filters clearly show a bipolar nebula with
a dark lane between the two lobes (Fig.\,\ref{lkha233_disk}). This lane is 
already indicated by the \citet{Cor98} finding that the
redshifted part of the bipolar jet is not visible for the first 0\farcs7 from
its origin, which is probably due to circumstellar dust occulting the receding
flow. \citet{Per04} have observed the dark lane on a larger spatial scale
(FWHM = 0\farcs27) using polarimetry in the near-infrared. The bipolar
structure seen in Fig.\,\ref{lkha233_disk} (left panel), together with the
`spikes' mentioned above, can thus be interpreted as a conical cavity cleared
by the jet, as has also been suggested by \citet{Per04}. The optical emission
comes from the inner edge of this cone that is illuminated by the star. An
important consequence of this is that the optical and near-infrared brightness
maximum does probably not coincide with the star itself, but with a region on
the surface of the cone pointing towards us. This is supported by the finding
that this maximum appears elongated and does not resemble a stellar PSF
(Fig.\,\ref{lkha233_disk}, right panel). The interpretation of the brightness
maximum as a scattering surface can also explain the extended structure found
by \citet{Lei93} that was interpreted as a halo around the star in that
work. From the present data, it is not clear if the `dark lane' is a
circumstellar disk seen close to edge-on or a dust torus around the star. If
the former interpretation is true, LkH$\alpha$~233 is one of the few young
stellar objects where a disk has been directly observed. This disk, however,
would have an apparent radius of about 1000~AU (adopting a size of the lane of
approximately 1\farcs1 at a distance to LkH$\alpha$ 233 of 880 pc, see
Fig.\,\ref{lkha233_disk}, left panel) and so it would be much larger than
disks around classical T~Tauri stars. The location of the bipolar jet is
indicated in Fig.\,\ref{lkha233_disk}. It is, however, not visible there
because the continuum emission from the reflection nebula overwhelms the line
emission from the jet. 

\begin{figure*}
\centering
\includegraphics[width=14cm]{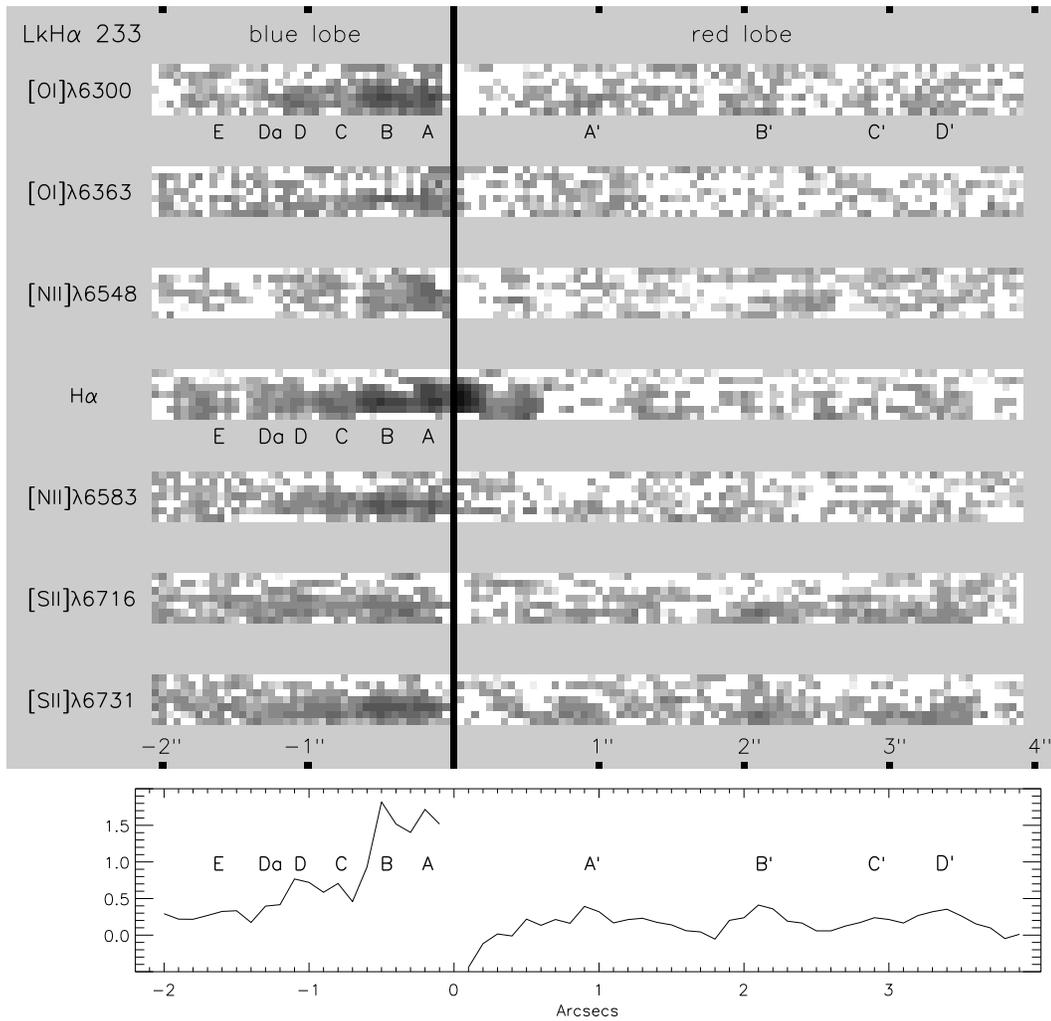}
\caption{\label{lkha233_stis} Images of the LkH$\alpha$~233 jet in different
  emission lines, reconstructed from HST/STIS spectra, integrated over the
  full radial velocity range of the lines. The thick solid line marks the peak
  of the optical brightness distribution in the reflection nebula surrounding
  the star. The redshifted outflow lobe (PA~=~68$^{\circ}$) is on the right
  side of the image. Each panel has a spatial width of $\approx 0\farcs5$. The
  intensity is displayed logarithmically from $5\times10^{-17}$ to
  $3\times10^{-14}$ erg\,s$^{-1}$\,cm$^{-2}$\,arcsec$^{-2}$. For H$\alpha$ in
  the redshifted lobe, the signal is heavily affected by contributions from
  the disk/envelope in the first 0\farcs5 from the central position. The
  bottom panel shows the total flux distribution coadding all forbidden lines
  along the jet.} 
\end{figure*}

\subsection{Jet structure}
\label{morph}

Spectroscopic data obtained with HST/STIS in the way described in Sect.\,
\ref{obs}, allow in principle to construct high angular resolution images
of the flow in distinct velocity intervals. This is done by selecting the same
velocity bin for each line in the S1...S7 spectra and aggregating the
selected seven pixel columns in a 2D image. The result is analogous to what
would be obtained with a high angular resolution integral field unit. This has
been done successfully in the past for the jets from the classical T Tauri
stars (CTTSs) DG Tau by \cite{Bac00} and RW Aur by \cite{Woi02}. In the
present case, however, the signal-to-noise ratio of the individual spectra of
LkH$\alpha$~233 is rather low. Therefore we could not reach a similar level of 
detail for this jet, but we aimed at constructing high angular
resolution maps integrated spectrally, by coadding (after subtraction of the
background) the column pixel values in each spectral channel.  The resulting
fluxes were strong enough to produce reconstructed images in each of the
emission lines integrated over the whole radial velocity range covered by the
lines with a sampling of 0\farcs05/pixel. Note that, since the STIS slit was
moved from NW to SE, the numbering of the slit positions goes from S1 at the
bottom pixel row of each panel to S7 at the top pixel row. The images in the
emission lines included in the spectra are shown in
Fig.\,\ref{lkha233_stis}. The lines are integrated over a velocity range of
$\approx [0, 225]\, \mathrm{km}\,\mathrm{s}^{-1}$ for the redshifted lobe and
$\approx [0, -225]\, \mathrm{km}\,\mathrm{s}^{-1}$ for the blueshifted lobe,
and the intensity varies logarithmically from $5\times10^{-17}$ to $3\times
10^{-14}$ erg\,s$^{-1}$\,cm$^{-2}$\,arcsec$^{-2}$. To make positions of the
jet knots more evident, we show in the bottom panel the total flux
distrubution coadding all forbidden lines along the jet. 

The redshifted lobe of the jet (PA~=~68$^{\circ}$), displayed in the right
part of the image, is longer, but considerably fainter than the blueshifted
outflow that occupies the lower part. The thick solid line marks the peak of
the optical brightness distribution, which, as mentioned above, represents a
bright point of the reflection nebula, and it is slightly displaced from the
position of the star. This also caused the jet not to be centred on each
individual panel, as explained above. 

The blueshifted jet lobe is clearly visible in the STIS data within its first
two arcseconds. Five jet knots (labelled A...E in Figs.\,\ref{lkha233_stis}
and \ref{lkha233_exc}) can be resolved in this region close to the star, with
possibly an additional faint knot (that we labelled Da) between D and E. The
knots are seen more clearly in [OI]\,$\lambda$6000, H$\alpha$, and
[SII]\,$\lambda$6731, but they can also be distinguished in [NII]\,$\lambda$6583,
where they appear to be more concentrated toward the jet axis.  This is
expected, since the [NII] lines are a tracer of higher excitation which, in
turn, is expected to be found in the axial region, where the jet shocks are
more powerful. If an arbitrary jet inclination of $45^{\circ}$ is assumed, the
knot separations from the presumed position of the jet source are 200, 600,
1100, 1300, and 2100~AU, respectively. There is general consensus on the fact
that bright knots trace the radiative cooling regions behind shock fronts,
which form in the beam as a consequence of pulsation of the ejection mechanism
at the jet base and/or internal instabilities during the propagation. The
distance of the jet from Earth, however, does not allow the STIS resolution to
ascertain whether the knots are actually bow-shaped, as seen in some other YSO
jets. 
For the redshifted counter-jet, four jet knots (labelled $A'...D'$ in
Fig.\,\ref{lkha233_stis}) are visible in the first $4''$ of the flow in
[OI]\,$\lambda$6300 and [SII]\,$\lambda\lambda$6716,6731. The S/N ratio is, 
however, quite low here. We note that the knots in the two lobes are not 
placed symmetrically with respect to the source position, as one would expect 
if, e.g. the knots were due to instabilities in the accretion/ejection
events. 
Similar asymmetries are found in many other bipolar jets, such as e.g., the
flow from RW Aur \citep{Woi02}, but an explanation for these asymmetries is
still lacking. Regarding the map in the H$\alpha$ line, we note that  the
emission of the jet was mixed with scattered light contributions from the disk
and the envelope. We thus used the signal from $S6$ and $S7$, which contains
no jet emission, to attempt a subtraction of these contributions. However, the
result should be taken with care, especially within the first 0\farcs5 of the
redshifted lobe. Here the signal appears to be heavily affected by the
contributions from the disk/envelope and may not represent structures within
the jet. 

The same jet has been observed at much coarser spatial resolution by
\cite{Cor98}, who spectrally resolved several velocity components and
determined corresponding electron densities,  as shown in their
Fig.\,2. Taking the difference in spatial resolution into account, their
results are in good agreement with our derivations. In Sect. \ref{disc} we
will compare the images shown in Fig.\,\ref{lkha233_stis} with analogous data
resulting from high angular resolution analyses of the base of CTTS jets from
other sources. 

\subsection{Physical quantities along the flow}
\label{phys}

\begin{figure}
\centering
\includegraphics[width=9cm]{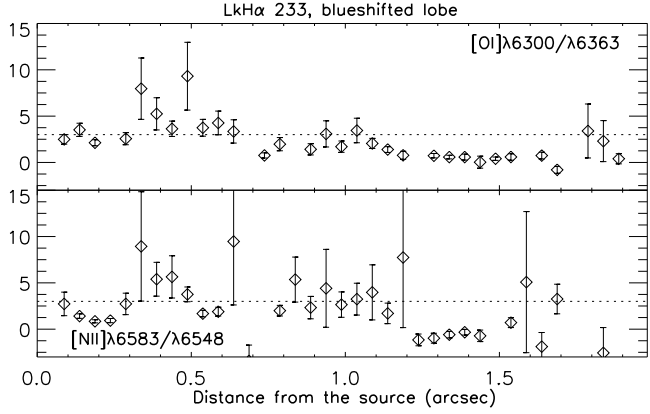}
\caption{\label{ratio1} Preliminary analysis of the ratios of selected
  forbidden lines along the blueshifted lobe of LkH$\alpha$~233, integrated
  across the full velocity range and across the jet transverse
  section. Distances are from the brightness maximum of the circumstellar
  reflection nebula, here referred to as `source'. Top panel:
  [OI]\,$\lambda$6300/$\lambda$6363 with its error, compared with the
  theoretical ratio of Einstein's A coefficients (dotted line). Bottom panel:
  same as above, for the [NII]\,$\lambda$6583/$\lambda$6548 ratio. In both
  panels discrepancies begin to be more evident from 1\farcs2 from the
  source.} 
\end{figure}

Thanks to YSO jets possessing  a rich spectrum of emission lines,
one can determine many interesting physical quantities associated with the
emitting gas. Besides the  accurate measurements of the radial velocity
through Doppler shifts, one can rely upon recently developed spectral
diagnostic techniques to find, for example, the ionisation fraction, the total
density, and the jet mass flux \citep{Bac99a, Har94}. 

To find values of these quantities in the LkH$\alpha$ 233 jet, we follow a
procedure described in \cite{Bac99a}, and subsequently refined and extended in
\cite{Pod06}, usually referred to as the {\em `BE technique'}, which uses a
combination of the  forbidden  lines from oxygen, sulphur, and nitrogen. These
lines, excited collisionally, can be considered optically thin in the density
and temperature regimes found in stellar jets and are easily
modeled. The H$\alpha$ emission, albeit very strong, is not used in the
diagnostics, because it is often contaminated by the reflection nebulosity
around the star, it can suffer from absorption in  high density zones, and it
is produced by both collisional excitation and radiative recombination in
regions not spatially coincident with the emission region of the  forbidden
lines \citep[see][]{Bac99a}. 

\subsubsection{Preliminary analysis of the line ratios}
\label{phys1}

Prior to the diagnostic analysis, the spectra were carefully examined, and it
was found that the redshifted lobe of the jet is too faint in our images for
the diagnostics to be applied at high angular resolution with a reasonable
level of confidence. Thus for this lobe we integrated the signal spectrally
and spatially in each of the four most prominent knots, and we applied the
diagnostics to each knot as a whole (see Table 1). 

For the brighter blueshifted lobe, the S/N was also not sufficient to perform
a 2D diagnostic analysis. Therefore, we integrated spectrally and across the
jet, and the  high angular resolution information was only retained for the
direction along the flow. For the blue-shifted lobe, we then formed various
ratios of the forbidden lines to check for the presence of evident problems
of S/N, that may render these lines unusable in certain
locations. The ratios [OI]\,$\lambda$6300/$\lambda$6363 and
[NII]\,$\lambda$6583/$\lambda$6548 should equal the ratio of
Einstein's A coefficients for spontaneous emission, which is about 3 in both
cases. Departure from this value can give an idea of the regions of the jet
where the  diagnostics will be less reliable due to poor S/N. The comparison
is shown in Fig.\,\ref{ratio1} with the theoretical value, and
the line fluxes were integrated across the jet and in radial velocity, with
errors propagated quadratically. 

The comparison shows that the observed points are fairly close to the
theoretical line until about $1\farcs2$ from the source, corresponding to the
tip of knot D in Fig\,\ref{lkha233_stis}. Beyond 1\farcs2 and up to 1\farcs7,
corresponding to a faint region of the jet, the observed ratio deviates from
the expected value. 
The ratio of oxygen lines is acceptable again at the end of the jet at the
location of knot E, while for nitrogen lines the ratio is still far from the
expected one at this position. Our test implies  that the diagnostic results
for the regions of the flow  farther from the source than knot D should be
taken with caution, especially for the ionisation fraction, which is tightly
correlated with the [NII] measurements. Since it is likely that these problems
are caused by the very low S/N of the faintest lines in each of these pairs,
we decided to use the flux of the brightest one and the theoretical ratio for
our subsequent excitation analysis. 

\begin{figure}
\centering
\includegraphics[width=9cm]{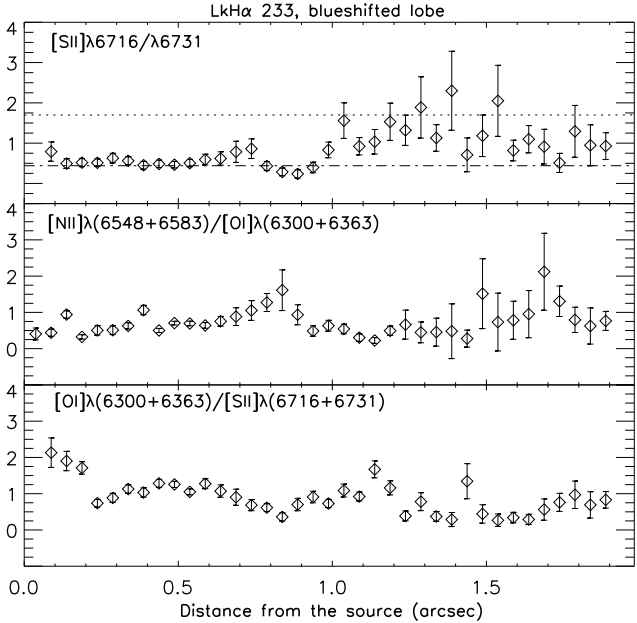}
\caption{\label{ratio2} Preliminary analysis of the line ratios used for the
  diagnostics  along the blueshifted lobe of LkH$\alpha$~233. Fluxes are
  integrated across velocity and spatially across the jet transverse
  section. Distances are from the brightness maximum of the circumstellar
  reflection nebula. Top panel: [SII]\,$\lambda$6716/$\lambda$6731 ratio with
  its measurement error. The theoretical value  in the low  density limit
  (dotted line) and high density limit (dot-dashed line), calculated for $T_\mathrm{e}
  = 10^4$ K are superimposed. Middle panel: ratio
  [NII]\,$\lambda$(6548+6583)/[OI]\,$\lambda$(6300+6363), with its error. This
  ratio is very sensitive to the hydrogen ionisation fraction (see
  text). Bottom panel: variation of the ratio
  [OI]\,$\lambda$(6300+6363)/[SII]\,$\lambda$(6716+6731), which is sensitive to
  both the electron temperature and the ionisation fraction of the emitting
  gas.} 
\end{figure}

The top panel of Fig.\,\ref{ratio2} illustrates the values of the ratio
[SII]$\lambda$6716/$\lambda$6731. From this ratio one can measure the electron
density along the flow. The theoretical value of the ratio calculated for $T_\mathrm{e}
= 10^4$ K in both the low and high density limits are superimposed on the
datapoints. The critical density
for this ratio (corresponding to the high density limit) is $2.5\times10^4$
cm$^{-3}$, while  $n_\mathrm{e}$ = 50  cm$^{-3}$ is the minimum value for the
subsequent diagnostics to be applicable. From the plot, one can argue that the
electron density along this jet is very high along the first 1\arcsec\ from the
source, while it clearly drops at greater distances, despite the larger
measurement errors. Where the measured [SII]\,$\lambda$6716/$\lambda$6731 values
are outside the indicated range (because of poor S/N), the data can only
provide upper/lower limits to the electron density, indicated in
the following plots with arrows. 

In Fig.\,\ref{ratio2}, we also report the observed value of the ratios
[NII]\,$\lambda$(6548+6583)/[OI]\,$\lambda$(6300+6363) (hereafter referred to as
[NII]/[OI], middle panel) and
[OI]\,$\lambda$(6300+6363)/[SII]\,$\lambda$(6716+6731) (hereafter [OI]/[SII],
bottom panel), integrated in velocity and across the jet cross section, with
errors propagated quadratically. As shown in \cite{Bac99a}, the  [NII]/[OI]
ratio only depends weakly on temperature, but it is a good indicator of the
level of global ionisation in the flow, since the ionisation fractions of N
and O are tightly correlated to that of hydrogen through charge exchange
reactions. The plot shows a gradual increase in this ratio, and hence of
ionisation, going from the source to the position of knot C. Then the values
decline to a minimum in the region of faint flux around 1\farcs4 from the
source. The ratio then appears to increase substantially at the end of the
flow; however, this may not be real due to the aforementioned measurement 
problems for the [NII] line fluxes. The [OI]/[SII] ratio depends on both the
hydrogen ionisation fraction and the electron temperature in the gas, but is
correlated more tightly  with the latter. 
The values of the ratio first decrease from the source to a minimum at
0\farcs2, followed by a  gradual increase along the jet, with the highest
values reached at the position of knot C, and a minimum immediately downstream
of knot D. The values of this ratio, however, are defined better toward the
end of the flow with respect to the [NII]/[OI] ratio. 

\subsubsection{Results of spectral diagnostics}
\label{phys2} 

In this paragraph we present the results of the BE diagnostic procedure
described in \cite{Bac99a} applied  to our dataset. Basically, the various
quantities are obtained by comparing the observed line ratios with those
calculated by a code designed to predict the fluxes in forbidden lines
starting from a grid of values of electron density $n_\mathrm{e}$, hydrogen ionisation
fraction $x_\mathrm{e}$, and electron temperature $T_\mathrm{e}$. The electron density  is
obtained by inverting the ratio of sulphur lines, which is independent of
$x_\mathrm{e}$ and only weakly dependent on $T_\mathrm{e}$. Then, $x_\mathrm{e}$ and $T_\mathrm{e}$ are found by
combining and inverting the [OI]/[NII] and [OI]/[SII] ratios, also using the
retrieved $n_\mathrm{e}$. The calculation was performed with a numerical code developed
by one of us (U. Locatelli) to apply the technique to large datasets. For more
details on the diagnostic technique and on the underlying physical
assumptions, together with its limits of application, see \citet{Bac02} and
\citet{Pod06}. 

Due to the aforementioned problems of poor S/N, we performed the
diagnostic analysis on the ratios obtained integrating the spectral datacube
both in velocity and in the spatial direction across the jet. Thus high
angular resolution is retained only for the direction along the flow. We
stress that, in the present case, due to the problems of low S/N,
we only use the brightest lines of the [OI] and [NII] doublets to form the
observed ratios that will be compared with the theoretical ones in the
diagnostics (see Sect.~\ref{phys1}). 

\begin{figure}
\centering
\includegraphics[width=9cm]{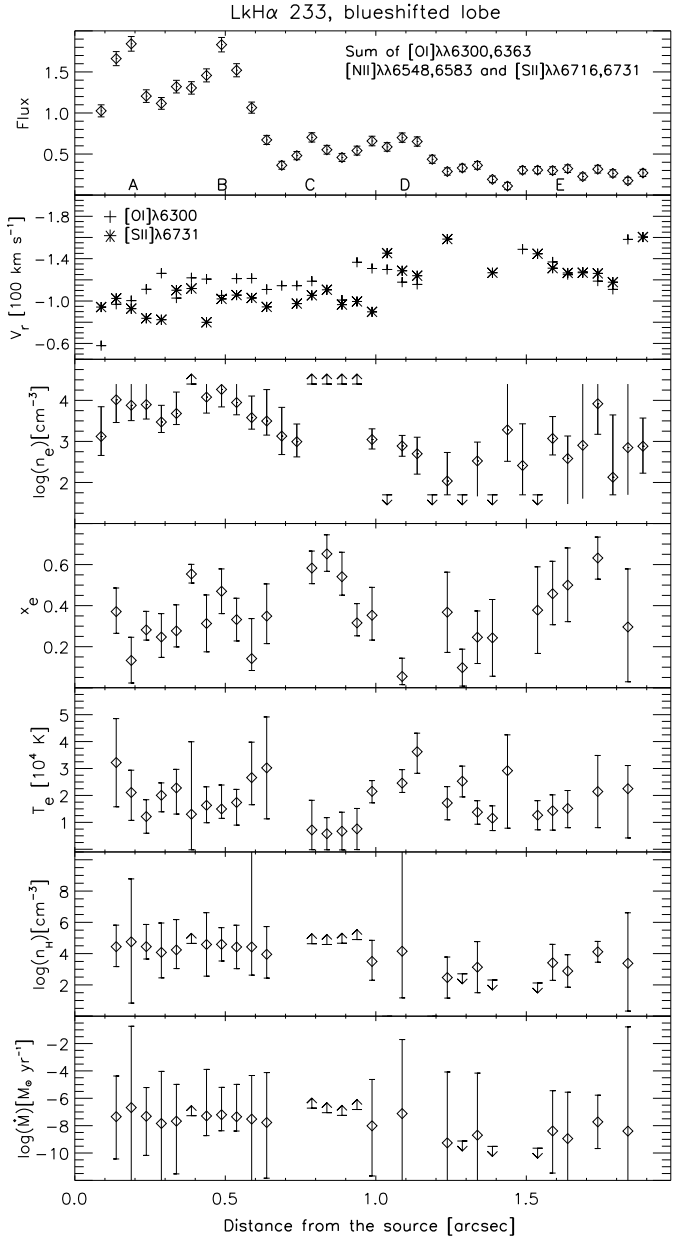}
\caption{\label{lkha233_exc} Physical properties of the emitting gas in the
  blueshifted lobe of the bipolar jet from LkH$\alpha$~233 as a function of
  distance from the brightness maximum of the circumstellar reflection
  nebula. All quantities (except the radial velocity) are derived after
  integration across wavelength and across the jet width. From top to bottom:
  intensity profile resulting from the sum of forbidden line fluxes, in units
  of $10^{-14}$ erg s$^{-1}$ cm$^{-2}$ arcsec$^{-1}$ (letters refer to the jet
  knots displayed in Fig.\ref{lkha233_stis}); radial velocity resulting from
  the analysis of the lines [OI]\,$\lambda$6300 and [SII]\,$\lambda$6731,
  integrated across the jet (errors range between 10 and 40 km s$^{-1}$, see
  text); electron density $n_\mathrm{e}$; hydrogen ionisation fraction $x_\mathrm{e}$; electron
  temperature $T_\mathrm{e}$; total density $n_\mathrm{H} = n_\mathrm{e}/x_\mathrm{e}$; mass loss rate
  $\dot{M}_\mathrm{J}$.} 
\end{figure}

The results of the diagnostics are collected in Fig.\,\ref{lkha233_exc}. The
figure, in the first two panels, also illustrates the intensity profile of the
emission lines and the radial velocity of the material at each position along
the jet, to give a positional and kinematical reference for the values of the
thermal properties shown in the other panels. 


The line flux distribution along the flow (top panel) is derived from the sum
of the spectroscopic images in  [OI]\,$\lambda\lambda$6000,6363,
[NII]\,$\lambda\lambda$6548,6583, and [SII]\,$\lambda\lambda$6716,6731, and
subsequent integration of the combined image across the jet axis. As mentioned
above, this step was necessary to reach a sufficient S/N ratio for
application of the diagnostics. 


The information about the jet structure is complemented by determining
of the radial velocity of each feature in the [SII]\,$\lambda$6731  and the
[OI]\,$\lambda$6300 lines (second panel from top). This was obtained by summing
the spectra in the direction transverse to the jet, which is coadding the
individual $S1$...$S7$ and then performing Gaussian fits to the integrated
(spatially) line profile in each pixel row. This yielded peak radial
velocities versus distance from the source, which were then transformed to
systemic velocities assuming the heliocentric radial velocity of
LkH$\alpha$~233 to be $-17.2\,\mathrm{km}\,\mathrm{s}^{-1}$ (see
Sect.\,\ref{intro}). 

The velocities along the line of sight vary between $-60$~km~s$^{-1}$ close to
the star and $-160$ km s$^{-1}$ at the end of the observed portion of the
jet. Errors are not reported on each point to keep the plot readable, but
these were accurately determined from the Gaussian fit and vary between 10
and 40 km s$^{-1}$ for both lines, being greater toward the end of the flow,
which is fainter. For the redshifted lobe, no precise Gaussian fit was possible
due to the low S/N. A rough estimate of the line shifts, however, 
indicated velocities of the plasma of about the same amplitude as in
the blue lobe. 

The velocity determinations indicate a gradual increase in velocity with
distance in the examined jet portion. This increase is, however, not
continuous, as the velocity is fairly constant along the first 0\farcs9 of the
jet (the brightest jet portion), and then presents sudden jumps at knots D and
E. This probably indicates that the last two knots move into a more rarefied
external medium. Internally to almost all of the knots, one can marginally
distinguish a local decrease in the radial velocity across each knot, revealed
in both lines. For an interpretation of this effect, see Sect.\,\ref{disc}. 

The lower panels of Fig.\,\ref{lkha233_exc} illustrate the thermal properties
in the blueshifted lobe of the jet, resulting from the application of the BE
technique. 


The third panel from top describes the variation in the electron density,
$n_\mathrm{e}$, along the jet, as derived from the inversion of the ratio of the two
sulphur lines for an assumed temperature of 10$^4$ K. The reported errors on
$n_\mathrm{e}$ come mainly from the propagation of the S/N uncertainty of the [SII]
lines, with a smaller contribution from the {\em a priori} uncertainty of
$T_\mathrm{e}$, although the line ratio is only weakly sensitive to this
parameter. `Upward' arrows are lower limits, equal to the critical density, in
the positions where the ratio reached its saturation value. Similarly,
`downward' arrows indicate positions where the electron density is lower than
50 cm$^{-3}$ (the limit of applicability of the BE technique). As already
argued from the inspection of Fig.\,\ref{ratio2}, the electron density is
quite  high in the first portion of the flow. Its value scatters around 10$^4$
cm$^{-3}$ in the first 0\farcs5 and then gently decreases down to 10$^3$
cm$^{-3}$ at 0\farcs75, corresponding to the base of knot C. It is then
apparently extremely high inside this knot, even though the corresponding
line brightness is not exceedingly large. The region downstream of knot C
presents a different character; since it is much less dense in free
electrons. Here
$n_\mathrm{e}$ scatters between 500 to 1000 cm$^{-3}$, apart from those positions where
the lower limit is reached, and an isolated point of higher density in knot
E. Overall, the situation appears to be consistent with the jet experiencing
less confinement by the surrounding medium downstream of knot C. 


In the fourth and fifth panels from top, we illustrate the results of the BE
diagnostics for the hydrogen ionisation fraction $x_\mathrm{e}$ and the electron
temperature $T_\mathrm{e}$ along the blueshifted jet. 
In determining these quantities, great attention has been devoted to a
realistic estimate of the errors, as there are a number of different sources
of uncertainty in this case. First of all, the diagnostic uses ratios of lines
produced by different elements, therefore an assumption of the set of
elemental abundances is necessary. This aspect is extensively analysed in
\cite{Pod06}, where the authors conclude that solar abundances are not
adequate in all cases, but instead the abundance set appropriate for the
region under study should be chosen. Unfortunately, no determination of
abundances exists in the literature for the Lacerta Cloud, so we have adopted
the abundance set of Orion by \citet{Est04}, assuming that the two regions
have been exposed to the same enrichment processes \citep[see][]{Neu06}. 
Another source of error comes from the uncertainty in $n_\mathrm{e}$, which is an input
of the diagnostic calculation. We then have to consider the measurement error
of the ratios [NII]/[OI] and [SII]/[OI], as displayed in Fig.\,\ref{ratio2},
which includes both the S/N accuracy and the effect of variations in
extinction along the jet beam that are unaccounted for. 

All these sources of error were considered and their effect on the final 
determination of $x_\mathrm{e}$ and $T_\mathrm{e}$ was found {\em a posteriori} by running 
the code at the extremes of the range of variations
in the input quantities. The resulting uncertainties are displayed as error
bars in the panels. These are limited to 15--20\% in the bright part of the
jet, but they rise up to 50--60\%  on average in the faint final
section. Missing points are due  to the fact that in those positions the
values of the ratios fall out of the range of validity of the technique. This
basically occurs in regions of poor S/N in the [NII] lines, but see
\cite{Pod06} for a more detailed discussion of this point. It has also to be
kept in mind that the positions where the electron density was only determined
as a lower  limit are critical, and the results for $x_\mathrm{e}$ and $T_\mathrm{e}$ must be
taken with caution in these regions. A better determination of the electron
density in these positions would be desirable, but this would require the use
of pairs of lines of  higher critical density not included in our STIS
setting. To overcome this problem, \cite{Har07} have recently attempted a
simultaneous inversion of all the line ratios available from similar STIS
spectra of HH~30 in the high density regions toward the star. The results for
HH~30 are encouraging, but the mathematical complexity involved in implementing
our code along the same lines does not seem justified by the very limited
number of positions in this jet where the [SII] ratio gives poor information. 

Despite the many sources of error discussed above, our diagnostics allowed us
to establish with reasonable accuracy both the value and the variation in the
ionisation fraction and the electron temperature in most of locations along
the jet. In the first bright portion $x_\mathrm{e}$ appears on average to increase from
about 0.2 in knot A up to 0.6 in knot C, where the jet shows the maximum
ionisation. Note that the latter corresponds to a lower limit determination of
the density, which may lead to a possible overestimate of the ionisation
fraction. An inspection of Fig.\,\ref{ratio2}, however, reveals that the
[NII]/[OI] ratio itself presents a peak in that position, which lead us to
believe that the enhancement of ionisation in this knot is real. Downstream of
knot C $x_\mathrm{e}$ falls again to values around 0.3, but it clearly rises again in
knot E up to $0.4-0.5$. While the variations are moderate, they are
nevertheless evident, and we note no clear association between peak intensity
and peak ionisation inside each knot. 

The electron temperature $T_\mathrm{e}$ in the jet ranges between 1 and $3\times10^4$
K, and its pattern of variation is complex, too. The value of $T_\mathrm{e}$ shows a
marked decrease 
in the first $0\farcs 3$ inside knot A, down to 10$^4$ K at 0\farcs25, and it
then turns up again to $3\times10^4$ K at the end of knot B. The temperature
shows a minimum in knot C, but here again the decrease to $7000-8000$ K may be
artificially caused by the poor determination of the electron density in this
knot. In knot D there are less positions in which the diagnostic is
applicable. Here, $T_\mathrm{e}$ first appears to increase again to $3-4\times10^4$ K
at 1\farcs1 from the source and then to decrease to $1.5-2\times10^4$ K in 
knots Da and in the first part of knot~E. There is a final local peak of the
temperature to $2.2\times10^4$~K  toward the second half of knot E. 


Once $x_\mathrm{e}$ and $n_\mathrm{e}$ are determined, a rough estimate of the average total
hydrogen density $n_{\mathrm{H}}$ can then be derived as $n_{\mathrm{H}} =
n_{\mathrm{e}} / x_{\mathrm{e}}$ (errors are propagated quadratically from
these quantities). Note, however, that this simple calculation likely
overestimates the actual average densities in the jet by a factor $3-5$,
because the determination is biased toward the regions of higher line
brightness, which in turn are those of higher compression after a shock
front. Nevertheless, it is extremely useful to have at least an approximate
value for this parameter, because the total density enters all the
relationships governing the dynamics of the jet. In particular, it is
extremely important to know this parameter when attempting a comparison
between the analytical and numerical model predictions and the actual
observations. The jet appears to be quite dense in the initial portion, as the
hydrogen total density  varies between  10$^4$  and 10$^5$ cm$^{-3}$ in knots
A--D. Of course, the lower limits found for $n_\mathrm{e}$ affect density estimations
inside knot C. Downstream of knot D, $n_{\mathrm{H}}$ presents lower values,
scattering on average around $5\times10^4$ cm$^{-3}$. Overall, the jet from
this source appears to be one of the densest studied so far with the BE
diagnostics. 

\begin{table}
\centering
\caption{\label{redlobe} Physical quantities in the redshifted lobe of the jet
  from LkH$\alpha$ 233, averaged  spatially and spectrally over each jet knots
  (see text).} 
\begin{tabular}{l@{\hspace{2mm}}c@{\hspace{2mm}}clc@{\hspace{2mm}}c@{}}
\toprule
Knot & Location & $\log(n_{\mathrm{e}})$ & $x_{\mathrm{e}}$ &
 $\log(n_\mathrm{H})$ & $\dot{M}_\mathrm{J}$  \\
 & ($''$) & (cm$^{-3}$) &  & (cm$^{-3}$) & (10$^{-7}$\,M$_{\sun}$\,yr$^{-1}$) \\ \midrule
$A'$  & $0.65-1.00$  & 4.06  & 0.12  & 4.98  & 3.7 \\
$B'$  & $1.85-2.50$ & 3.85  & 0.19  & 4.58  & 1.5 \\
$C'$  & $2.55-2.95$ & 3.94  & 0.47  & 4.26  & 0.7 \\
$D'$  & $3.00-3.45$ & 3.90  & 0.20  & 4.59  & 1.5 \\
\bottomrule
\end{tabular}
\end{table}

The last panel of  Fig.\,\ref{lkha233_exc} illustrates our estimate of the
mass flux in the jet. This is discussed in detail in the next section. 

We conclude this part by commenting briefly on the diagnostic results obtained
for the red lobe, from the data integrated spectrally and spatially in each
knot (see Table \ref{redlobe}). The excitation conditions are quite different
here from those found in the blue lobe, which is already indicated by the fact
that this counter-jet is more prominent in [SII]\,$\lambda\lambda 6716, 6731$
than in [OI]\,$\lambda 6300$, in contrast to the blueshifted lobe. From the
table one can see that $n_{\mathrm{e}}$, $x_{\mathrm{e}}$, and
$n_{\mathrm{H}}$ are all lower in the redshifted lobe than in the blueshifted
one. In particular $n_{\mathrm{e}} < n_{\mathrm{crit}}$ of the sulphur lines
for all the knots investigated here. The maximum electron density is found in
knot A$'$, where it is about 10$^4$ cm$^{-3}$. It then decreases in the other
knots, but it is never less than 8000 cm$^{-3}$. On the other hand, the
ionisation fraction is generally not more than $10-20$\%, if exception is made
for knot C$'$, which seems quite excited, reaching $x_\mathrm{e} \sim 0.5$. The
combination of $x_\mathrm{e}$ and $n_\mathrm{e}$ provides quite a high total density that
decreases from  10$^5$  cm$^{-3}$ at the base of the jet down to $\sim
1.8\times10^4$  cm$^{-3}$ in knot C$'$. The electron temperature is around
$1-2\times10^4$~K. Because of the low signal in [OI] lines, however, the
estimate is not as detailed as in the blue lobe, so values are not given in
the table. We instead attempted an estimate of the mass flux, which is
discussed in the following section.  

\section{Discussion}
\label{disc}

\subsection{Mass flux in the jet}
\label{mdot}

One of the most useful quantities  to know when comparing models with
observations in the case of jets is the mass flux in the outflow, $\dot{M}_\mathrm{J}$,
and its variation along the flow. For example, in the magnetohydrodynamic
(MHD) models proposed to explain the formation of YSOs, the ratio between
ejected and accreted matter around the star is fixed ($\dot{M}_\mathrm{J}/\dot{M}_{\rm
  acc} \sim 0.01 \div 0.1$). Moreover the knowledge of $\dot{M}_\mathrm{J}$ allows us
to estimate other important dynamical quantities like the linear and angular
momentum carried by the jet (adding the information on poloidal and, where
available, toroidal velocities). In turn, these estimates can clarify if the
jet is capable of accelerating coaxial outflows, to clear the stellar
environment and to inject turbulence in the cloud \citep{Pod06}, as well as to
extract the excess angular momentum from the disk/star system \citep[see
  e.g.][]{Woi05}. 

In principle we can determine the mass flux from the emission lines in two
ways, described in detail in \cite{Nis05}. In the first method, the mass flux
at each position  is estimated by combining the total density inferred from
diagnostic techniques with the values of the jet radius and velocities, $r_\mathrm{J}$
and $v_\mathrm{J}$, provided by the FWHM of the luminosity profile across the jet and
the kinematical analysis, respectively. This approach assumes that each
considered `slice' of the jet is uniformly filled at the density derived from
the diagnostics, which is biased toward more compressed regions. The second
method compares the observed luminosities of the forbidden lines to the one
calculated by an emission model, thus providing the mass of the emitting
gas. This method implicitly takes the filling factor into account, but is
affected by uncertainties in absolute calibrations, extinction, abundance, and
distance. In our case, the problems of an unknown filling factor is mitigated
by our observing the jet close to the source and at high angular resolution,
which limits the possibility of including large regions of non-emitting
material in the beam. On the other hand, extinction, elemental 
abundances, and distance are not well known in our case. Therefore, we prefer
to use the first method, keeping in mind  that the derived values of
$\dot{M}_\mathrm{J}$ have  to be considered as upper limits. Previous studies performed
on other jets by applying both methods show that the mass flux derived in this
way is overestimated by not more than 20\% when high angular resolution is
used \citep{Bac99b}. 

With our estimate of $n_{\mathrm{H}}$, the mass flux can be computed as
$
\dot{M}_\mathrm{J} = \mu m_p n_{\mathrm{H}}\,\pi\,r_\mathrm{J}^2\,v_\mathrm{J},
$
where $r_\mathrm{J}$ is the jet radius, $v_\mathrm{J}$ the full spatial velocity, $m_p$ 
the proton mass, and $\mu$ the  mean molecular weight, equal to $\sim$ 1.24 for
interstellar atomic gas in standard composition. 
In our case, the jet radius was estimated taking half the FWHM obtained from
Gaussian fits to the sum of all spectroscopic maps in forbidden emission
lines, which is nearly constant at $\approx 100$~AU between $d=0\farcs1$ and
0\farcs6. As the S/N is not high enough to derive reliable
results beyond $d =$ 0\farcs6, we assume the jet radius to be the same there
as closer to the origin, i.\,e. $r_\mathrm{J} = 50\,\mathrm{AU}$. The flow velocity
$v_\mathrm{J}$ is derived from fits to the coadded spectra $S1...S7$ in
[OI]\,$\lambda$6300 and [SII]\,$\lambda$6731, as mentioned above, and de-projected
with an arbitrary inclination angle of $i = 45^{\circ}$. The resulting mass
flux in each position of the blueshifted lobe is shown in the bottom panel of
Fig.\,\ref{lkha233_exc} (errors also include the uncertainties on  $v_\mathrm{J}$ and
$r_\mathrm{J}$ estimated from the Gaussian fit to the profiles of the emission
lines). The mass flux appears to be fairly constant at a value of about
$10^{-7}$ M$_{\sun}$ yr$^{-1}$ between the source and knot C, indicating that
the flow is well-collimated in this portion and that turbulent interaction
with the external medium, which may change the amount of transported mass, is
negligible. The mass flux then decreases by $1-2$ orders of magnitude in the
last part of the jet. This could be because the real jet radius
is actually larger than the one assumed. The fact that the jet undergoes
lateral expansion downstream of knot C is also supported by the marked change
of the other physical properties derived by the diagnostics. Alternatively, a
significant portion of the mass could have been ejected sideways by the bow
shocks in the working surfaces during the jet propagation, or $\dot{M}_\mathrm{J}$ may
be higher now than it was in the past. 

Regarding the redshifted lobe, for which our derivations come from data
integrated spatially and spectrally on each condensation, we can estimate the
mass flux averaged over each knot as was done, e.g., in
\cite{Pod06}. To this purpose, we assume a de-projected flow velocity of
170~km\,s$^{-1}$  and a jet diameter of 100~AU, similar to the blueshifted
lobe. The obtained values are reported in Table~1. The resulting mass flux in
the counter-jet seems to have the same order as in the blueshifted jet, with a
maximum for knot $A'$, where we obtain $\dot{M}_\mathrm{J} \sim 3.7\times10^{-7}
M_{\sun}$~yr$^{-1}$, and lower values in the following knots, but still
the same order of magnitude. 

Overall, the value of the observed mass flux in proximity to the star is
slightly above the upper boundary of the values obtained for the jets from T
Tauri stars. If the ratio of accreted-to-ejected mass flux is regulated as
suggested by standard MHD jet launching models, the results would
imply a mass accretion, $\dot{M}_{\rm acc}$, onto the star of the order of
several $10^{-6} M_{\sun}$ yr$^{-1}$. 

\subsection{Nature of the knots}

In the second panel of Fig.\,\ref{lkha233_exc}, internal to almost all of
the knots, one can marginally distinguish a local decrease in the velocity,
revealed in both the [OI] and [SII] lines. This is consistent with the knots
being mini-working surfaces, which should form inside the jet according to
some models of jet propagation, and are also directly imaged in resolution 
observations of T Tauri jets \citep{Ray96}. Each working surface is contained
within two shocks, one on the side of the jet source, the so-called Mach disk,
in which the high velocity material of the jet beam coming from the source is
decelerated upon entering the working surface. The other shock is on the
downstream side, the bow-shock, in which the material ahead and moving more 
slowly 
than the shock front, is accelerated upon entering the working surface. The
cooling region of the working surface is located between the two shocks and
can be assumed to extend about 10$^{15}$ cm, which is the same scale as 
the knots observed in this flow. It is therefore reasonable to interpret the
local variation in the velocity observed within each knot as a consequence of
the combined action of the two shocks constituting a working surface. In this
framework, the slight differences in the position of the velocity variations
in the two emission lines can be due to the peak emission region
of the two species behind a shock front being slightly displaced
\citep{Bac99a}. In conclusion, even if a bow-shaped morphology of the knots is
not directly identified in the image  of this flow, the velocity pattern
appears to point toward a similarity of the knot properties with those of the
TTS jets. 

\subsection{Comparison with flows from lower mass stars}
\label{mdot1}

These findings can be compared with studies of other jets from T Tauri stars
studied at high angular resolution and/or analysed with the same diagnostic
method. The \object{HH~30}, DG~Tau, and RW~Aur jets were observed at high
angular resolution with HST close to their sources ($d\le 1000\,\mathrm{AU}$)
\citep{Ray96, Bac99b, Bac00, Bac02, Woi02, Har07} and from the ground with the
CFHT using adaptive optics techniques \citep{Dou00}. These jets were then
analysed with the same diagnostic technique used here. Several other jets
(\object{HH~34}, \object{HH~1}, \object{HH~111}, \object{HH~46/47},
\object{HH~24}, \object{HH~83}, \object{HH~30}, \object{HL~Tau}, and
\object{Th~15-28}) were observed from the ground on much larger spatial scales
of 1000 to some 10000~AU from their parent stars and analysed with the BE
technique \citep{Bac99a, Nis05, Pod06}. In these cases the sources are either
deeply embedded protostars or T~Tauri stars. 

In the first $\approx 2000$~AU of the blueshifted LkH$\alpha$~233 jet, the
electron density $n_{\mathrm{e}}$ is remarkably higher than the one found in
jets at large distance from the star, but similar to the electron density
values found at the base of the jets observed at high angular resolution
\citep{Bac00, Bac02}. On larger scales $n_\mathrm{e}$ is expected to decrease due to
recombination and the widening of the flow. 

In the HH~30 jet, analysed at high spatial resolution, the ionisation is lower
on the whole than in LkH$\alpha$~233. Likewise, it is rising in the first part
of the flow and reaches its first maximum at $d \approx 300$~AU (in contrast
to $d \approx 500$~AU for LkH$\alpha$~233). In the same region the temperature
decreases steeply from $2\times10^4$ to less than $10^4$~K. Similar behaviour
is also seen in the DG Tau and RW Aur jets observed with HST at high angular
resolution (\citealt{Bac02}, Bacciotti et al. in preparation, Melnikov et
al. in preparation), thus this appears to be a common phenomenon. Several
studies have indeed recently tried to justify these trends in the framework of
jet acceleration models \citep[cf.][]{Sha02,Gar01}. Alternatively, knot~A
might trace the presence of a strong shock, where $T_{\mathrm{e}}$ is high and
$x_{\mathrm{e}}$ first increases to reach a plateau and then `freezes' and
slowly recombines along the flow \citep[cf.][]{Bac99a}. For the other jets
studied on larger angular scales, the ionisation is of the same order of
magnitude as for LkH$\alpha$~233. Re-ionisation events are found in a number
of these outflows, but they occur on a much larger spatial scale than in the
jet studied here. 

The total density in the LkH$\alpha$~233 jet, derived by combining the
ionisation fraction and electron density, is quite high, at least in its
initial part, reaching 10$^5$ cm$^{-3}$. Overall, the jet from this source
appears to be the densest studied so far with the BE diagnostics. 

The electron temperature $T_{\mathrm{e}}$ in the LkH$\alpha$~233 jet is of the
same order of magnitude as for the other jets. Peaks of $T_{\mathrm{e}}$ along
the jet, as in the blueshifted lobe of the LkH$\alpha$~233 jet, can also be
seen in the outflows of HH~46/47, HH~24~C, and HL~Tau, but again at much
larger separations from their respective sources. Finally, the observed flow
velocity is not very different from that of T Tauri jets. 

The mass loss rate is another fundamental parameter for testing the similarity
of 
the physics in YSO outflows. In the case of LkH$\alpha$~233 the mass loss rate
is in general larger by up to one order of magnitude than for the jets
observed on extended scales. It is, however, only slightly higher than the
values found for other jets from lower mass stars studied at high angular
resolution close to the source. 
In conclusion, the physical properties of this jet appear similar to those of
classical T Tauri stars, apart from an enhancement of the total density. 

\citet{Gar06} have determined accretion rates for a number of Herbig stars
using a well-established correlation between $L($Br$_\gamma)$ and the
accretion luminosity $L_\mathrm{acc}$. For a set of 36 Herbig stars, they found that
80\% are accreting matter at rates of $3\times10^{-9}\lesssim
\dot{M}_\mathrm{acc}\lesssim 10^{-6}M_{\sun}$ yr$^{-1}$. If we adopt a mass flux for
LkH$\alpha$~233 of about 10\% of its mass accretion rate (according to MHD
models), then the $\dot{M}_\mathrm{acc}\sim 10^{-6}M_{\sun}$ yr$^{-1}$ (taking a
smaller fraction of ejected to accreted matter would lead to an even higher 
value of $\dot{M}_\mathrm{acc}$). This would make LkH$\alpha$~233 one of the
strongest accretors among the Herbig stars. \citet{Gar06} also found that the
$\dot{M}_\mathrm{acc}$ distribution of Herbig stars is roughly consistent with the
prediction of a $\dot{M}_\mathrm{acc} \propto M_\star^{1.8}$ relation derived for
subsolar mass objects and noticed a deficit of very strong accretors with
rates higher than $10^{-7}M_{\sun}$ yr$^{-1}$. They suggested that this
deficit could be due to an ``aging''-effect, since the accretion rate is
expected to decrease with time \citep{Hart98}. From this point of view
LkH$\alpha$~233 would be a very young Herbig star with a very high accretion
rate in which the jet carries away an excess of angular momentum. 

\section{Conclusions}
\label{concl}

We have analysed  images and spectra of the immediate environment of the
Herbig Ae/Be star LkH$\alpha$~233 taken in the visible range with the Hubble
Space Telescope, which provides unprecedented high spatial resolution (0\farcs1
in the optical). The observed brightness distribution in archival HST images
is probably from the surface of a cone; i.\,e. we do not see the source
directly. We infer that a disk or dusty torus obscures the star. The
brightness of the nebula does not allow one to see the jet in filtered
narrowband ($\Delta\lambda\approx 50$~\AA) images, but it could be imaged in
2D maps derived from STIS data, reconstructing  images of the flow in the
various lines integrated in velocity from a set of seven spectra taken
by stepping the slit in adjacent positions. 

In this way, we have resolved fine structure in the distant bipolar jet from
LkH$\alpha$~233 (880 pc), in both the blueshifted and redshifted lobes. Within
its first arcseconds ($\approx 2000$~AU), the jet knots appear circular or
elliptical at HST resolution, although the kinematical analysis reveals that
they probably trace mini-working surfaces, as is common in stellar jets. 

We have analysed the spectroscopic maps with a diagnostic code designed to
adapt the so-called BE technique \citep{Bac99a} to large datasets. The low
values of S/N of the spectra forced us to integrate the maps across the jet
width in the blue-shifted lobe, and across the extension of each knot in the
red lobe, in order to have acceptable inputs for the diagnostic technique. The
results are displayed in Fig.\,\ref{lkha233_exc} and Table \ref{redlobe},
respectively.  High angular resolution is retained along the jet axis in the
blueshifted lobe. 

The electron density in the first arcsecond of the blueshifted lobe is about 
10$^4$ cm$^{-3}$, close to the critical density for [SII] emission and, 
in a few positions, is above the critical value. The electron density
decreases along the flow by slightly more than one order of magnitude,
reaching values of about $800-1000$ cm$^{-3}$ at the position of the last
visible knot (2$''$ from the source). 

The hydrogen ionisation fraction in the same lobe varies between 0.2 and
0.6. It gently rises on the first 500~AU of the flow, like other jets from
young stellar objects studied at high angular resolution; but instead of
reaching a plateau followed by slow decay, it goes through apparent episodes of
re-ionisation along the flow. 

The electron temperature $T_\mathrm{e}$ found with the diagnostics varies between 1
and $3\times10^4$ K, in agreement with the kind of lines observed. The jet
knot A, located closest to the origin, is correlated with a maximum of the
electron temperature  and might represent a strong shock. In the region of jet
knot C, there appears to be an anti-correlation between $T_{\mathrm{e}}$ and
ionisation. The nature of this anti-correlation is unclear, but the result for
the temperature here might have been affected by quenching of the [SII]
lines. 

Combining the values of $x_\mathrm{e}$ and $n_\mathrm{e}$ we obtained estimates of the total
density in this flow. The jet appears to be quite dense, especially close to
the star; and with its $n_\mathrm{H} \sim 10^5$ cm$^{-3}$, it turns out to be the
densest outflow studied with the BE technique so far. 

The mass flux $\dot{M_\mathrm{J}}$ is $\approx 10^{-7}$ M$_{{\sun}}$ yr$^{-1}$ in the
first arcsecond of the jet, while it decays by an order of magnitude further
out. These values are slightly higher than the mass-loss rate measured for the
jets from lower mass YSOs. This is the first determination of
a mass flux rate of a jet from a Herbig star obtained with the same method as
for T Tauri jets. 

In the redshifted counter-jet, electron density, ionisation, and mass density
are all lower than in the blueshifted jet and are consistent with a lower
degree of excitation. 

We have compared the outcomes of our study with the properties of jets from T
Tauri stars studied previously with the same diagnostics. The physical
properties of the jet from this HAeBe star appear totally analogous to those
in T Tauri jets, scaled toward higher densities and higher mass-loss
rates. This appears to be consistent with a similar accretion/ejection engine
in the two classes, with an enhanced rate of mass accretion onto HAeBes with
respect to TTSs. 

Overall, these results suggest the same type of `engine' drives the
accretion/ejection in intermediate and low mass stars as well as sub-stellar
objects. 

\begin{acknowledgements}
J.E., S.M., and J.W. acknowledge support from the Deutsches Zentrum f\"ur
Luft- und Raumfahrt under grant 50 OR 0009. The present work was supported in
part by the European Community Marie Curie Actions - Human Resource and
Mobility within the JETSET (Jet Simulations, Experiments, and Theory) network
under contract MRTN-CT-2004 005592. 
\end{acknowledgements}

\end{document}